\documentclass[aps,prd,superscriptaddress,notitlepage]{revtex4-1}

\usepackage{graphicx}
\usepackage{amsmath}
\usepackage{amssymb}
\usepackage{sistyle}
\usepackage{color}
\usepackage{mathtools}
\usepackage{slashed}

\newcommand{\bs}[1]{\mathbf{#1}}
\def\sign{\,\text{sign}\,}

\begin{document}

\title{Fermion self-energy in magnetized chirally asymmetric QED matter}

\author{D.O. Rybalka}

\affiliation{Department of Physics, Arizona State University, Tempe, Arizona 85287, USA}
\affiliation{Department of Physics,  Taras Shevchenko National Kiev University, 03022,  Kiev, Ukraine}

\date{\today }

\begin{abstract}
The fermion self-energy is calculated for a cold QED plasma with chiral chemical potential in a magnetic field. It is found that a
momentum shift parameter dynamically generated in such a plasma leads to a modification of the chiral magnetic effect current. It 
is argued that the momentum shift parameter can be relevant for the evolution of magnetic field in the chirally asymmetric primordial plasma in 
the early Universe.
\end{abstract}

\maketitle

\section{Introduction}

Recently the study of relativistic or pseudorelativistic systems in a magnetic field draw attention of researchers in areas
ranging from the quark-gluon plasma in heavy-ion collisions to primordial plasma in the early Universe and to quasiparticle excitations in
Dirac and Weyl semimetals. (For a review, see Ref.~\cite{Miransky:2015ava}.)
Probably the single most significant reason for the surge of interest in the study of these systems is connected
with the recognition that the chiral anomaly discovered long ago in high energy physics \cite{chiral-anomaly} is relevant for the transport
properties of relativistic fermion systems in a magnetic field.

In order to phenomenologically describe a chiral asymmetry in relativistic fermion matter,
it was proposed \cite{Fukushima:2008xe} to introduce a chiral
chemical potential $\mu_5$. This chemical potential couples to the difference between
the number of left- and right-handed fermions and enters the Lagrangian density through
the term $\mu_5\bar{\psi}\gamma^0\gamma^5\psi$. This produces a chiral asymmetry in magnetized relativistic matter and
leads to a non-dissipative electric current $\mathbf{j}=e^2\mathbf{B}\mu_5/(2\pi^2)$ in the presence
of an external magnetic field $\mathbf{B}$ \cite{Kharzeev:2007jp,Fukushima:2008xe}. This phenomenon is known in the literature as the chiral 
magnetic effect (CME). The similar effect takes place in the case of a plasma with nonzero chemical potential $\mu$, in which a non-dissipative axial current $\mathbf{j}_5=e^2\mathbf{B}\mu/(2\pi^2)$ is induced instead  \cite{Vilenkin:1980ft,Zhitnitsky}. This current makes fermions of different chiralities to travel in the opposite directions and so this phenomenon is called the chiral separation effect (CSE).

It was argued in Refs.~\cite{Zhitnitsky,Newman} that non-dissipative currents in magnetized relativistic matter are completely determined by
the topological currents induced only in the LLL and intimately connected with the chiral anomaly. This fact is
directly connected with the well known result that in a magnetic field the chiral anomaly is also generated only in the LLL
\cite{Ambjorn}. However, it was shown in Ref.~\cite{chiral-shift-1} that the normal ground state of such matter is characterized by a 
dynamically generated chiral shift parameter
$\Delta$. It enters the effective Lagrangian density through the following quadratic term: $\Delta \bar\psi \gamma^3 \gamma^5 \psi $.
The meaning of this parameter is most transparent in the chiral limit: it
determines a relative shift of the longitudinal momenta in the dispersion relations of opposite
chirality fermions, $k^{3}\to k^{3}\pm\Delta$, where the momentum $k^{3}$ is directed along
magnetic field \cite{chiral-shift-1}. 

The first studies of interaction effects on the chiral asymmetry of relativistic matter in a magnetic
field were done in Refs.~\cite{chiral-shift-1,chiral-shift-2,Fukushima} by using Nambu--Jona-Lasinio (NJL) like models
with local interaction. In particular, by using the Schwinger--Dyson (gap) equation, it was found
that the interaction unavoidably generates a chiral shift parameter $\Delta$ \cite{chiral-shift-1,chiral-shift-2} when
the fermion density is nonzero.
Furthermore, as shown in
Refs.~\cite{chiral-shift-1,chiral-shift-2,chiral-shift-3}, the chiral shift $\Delta$
is responsible for an additional contribution to the axial current. Taking into account that fermions in {\em all}
Landau levels, including those around the Fermi surface, are affected by $\Delta$,
the corresponding matter may have unusual transport and/or emission properties.
The dynamics responsible for the generation of the chiral shift parameter in the Nambu--Jona-Lasinio model was studied at finite
temperature in Ref.~\cite{chiral-shift-2}. It was shown that $\Delta$ is rather insensitive
to temperature when $T \ll \mu$, where $\mu$ is the chemical potential, and {\it increases} with $T$
when $T > \mu$. The first regime is appropriate for stellar matter, and the second one is realized in heavy
ion collisions.

Since the NJL model is nonrenormalizable and the chiral anomaly is intimately connected with ultraviolet divergencies, in order to reach a definite
conclusion about the presence or absence of higher-order radiative corrections to the axial current, one should consider them
in a renormalizable model. This was done in Refs.~\cite{Gorbar:2013upa,Xia:2014wla} in the limit of weak and strong magnetic field, respectively. Just like in the NJL model, it was shown that there are nonzero radiative corrections to the axial current in dense QED in a magnetic field. Also, chiral asymmetry was found to be produced not only by the lowest Landau level, but higher levels as well \cite{chiral-shift-QED}.

A nonzero chiral asymmetry could play an important role in the early Universe. As suggested recently 
in Refs.~\cite{Joyce:1997uy,Boyarsky:2011uy,Tashiro:2012mf,Manuel:2015zpa,Boyarsky:2015faa,Hirono:2015rla}, it may be connected to the primordial origin of large scale magnetic fields. It is plausible that pseudo-relativistic plasmas with nonzero $\mu_5$ could be also relevant for the recently 
discovered Dirac and Weyl semimetals. In the present paper, we study the relativistic matter with nonzero chiral chemical potential $\mu_5$ in a 
magnetic field at zero temperature. This study is a 
generalization to the case of a plasma with a chiral chemical potential of the previous investigations \cite{chiral-shift-1,chiral-shift-2}  of relativistic fermion matter in a magnetic field, where the usual chemical potential was used. 

The paper is organized as follows. In Sec.~\ref{sec:simpl}, we consider the Nambu-Jona-Lasinio model and find in the mean field
approximation that a momentum shift parameter is dynamically generated. However, since the results depend on the regularization scheme, more 
robust QED calculations in the one-loop approximation in the linear in magnetic field regime are performed in Sec.~\ref{sec:QED}. The structure 
of the Fermi surface and corrections to the dispersion relation are determined. The conclusion section provides a summary of the results.

\section{Momentum shift parameter in the Nambu-Jona-Lasinio model}
\label{sec:simpl}

In order to gain an insight into the role of interactions on the fermion self-energy we consider the NJL model of a chirally asymmetric plasma with a local interaction, assuming that the electromagnetic field can be considered at least qualitatively short-range due to screening effects. Thus we make use of the following Lagrangian density with a point-like effective interaction, whose strength is controlled by the constant $G_{\rm int}$:
\begin{equation}
	{\cal L} = \bar{\psi}i{\cal D}_\nu\gamma^\nu\psi + \mu_5\bar{\psi}\gamma^5\gamma^0\psi + \frac{G_{\rm int}}{2} \left[ (\bar{\psi}\psi)^2 + (\bar{\psi}i\gamma^5\psi)^2 \right],
\end{equation}
where the covariant derivative ${\cal D}_\nu = \partial_\nu - ie A^{\rm ext}_\nu$ includes external magnetic field in the Landau gauge $A_{\rm ext}^\mu=(0,0,xB,0)$. 
For the sake of simplicity, we set the electric chemical potential to zero, i.e., $\mu=0$.
Note that this Lagrangian possesses the $U(1)_L\times U(1)_R$ symmetry. 
The inverse free propagator has the form:
\begin{equation}
	iS^{-1}(u,u') = \left[ (i\partial_t+\mu_5^0 \gamma^5)\gamma^0 - (\bs{\pi}_\bot \cdot \bs{\gamma}) - \pi_3 \gamma^3\right] \delta^4(u-u') .
\end{equation}

Our ansatz for the inverse full propagator is similar to that in Ref.~\cite{chiral-shift-1} with $\mu$ replaced by $\mu_5\gamma^5$:
\begin{equation}
	iG^{-1}(u,u') = \left[
	(i\partial_t+\mu_5 \gamma^5)\gamma^0 - (\bs{\pi}_\bot \cdot \bs{\gamma}) - \pi_3 \gamma^3 + i \tilde{\mu} \gamma^1 \gamma^2 + \kappa \gamma^3
	\right] \delta^4(u-u'),
\label{eq:full-propagator-inverse}
\end{equation}
where for simplicity we consider the $m=0$ case. The physical meaning of the model parameters are easy to understand from the matrix structure of the corresponding terms: $\mu_5$ is a chiral chemical potential modified due to interactions, $\tilde{\mu}$ is the anomalous magnetic moment, $\kappa$ is the momentum shift along the magnetic field. The latter may be responsible for electric current.

The gap equation in the first order of interaction constant (one loop approximation) has the form \cite{chiral-shift-1}:
\begin{equation}
	iG^{-1}(u,u)-iS^{-1}(u,u) = 2 G_{\rm int} K(u,u).
\end{equation}

From the structure of the gap equation, it can be derived that $\tilde{\mu}=0$. To solve it we need to find the expression for the full propagator in the coincidence limit, which can be presented in the following form:
\begin{equation}
	\begin{aligned}
		K(u,u)
		=& \frac{\gamma^0\gamma^5 \mu_5}{4\pi^2 l^2} \sum_{n=0}^{[\mu_5^2/2|eB|]} \left[P_- + P_+ \theta(n-1) \right] \sqrt{1-2|eB|n/\mu_5^2} 
		\\
		&- \frac{\gamma^3}{8\pi^2 l^2} \sum_{n=0}^{\infty} \left[P_- + P_+ \theta(n-1) \right] \int dk_3  \frac{k_3-\kappa}{\sqrt{(k_3-\kappa)^2 + 2n|eB|}},
	\end{aligned}
\label{eq:full-propagator-in-one-point}
\end{equation}
where $l=1/\sqrt{|eB|}$ is a magnetic length, $P_\pm=\frac{1}{2}[1 \pm i s_\bot \gamma^1\gamma^2]$ is a projector on the spin states and $E=\sqrt{(k_3-\kappa)^2 + 2n|eB|}$ is the particle energy. Note that as expected energy dependence on momentum along $B$ is shifted by $\kappa$, whereas the transverse momentum is replaced by $\sqrt{2n|eB|}$, where $n$ is the Landau level quantum number. The momentum integral in Eq.~(\ref{eq:full-propagator-in-one-point}) is UV divergent and thus sensitive to a regularization used. We will apply the symmetric cutoff and proper-time regularization schemes to find the induced current in the system in order to extract the scheme independent features of the theory.

\subsection{Cutoff regularization}
First we apply the simplest regularization to the integral in Eq.~(\ref{eq:full-propagator-in-one-point}) by introducing an ultraviolet  momentum cutoff $\Lambda$. The full fermion propagator in the coincidence limit then takes the form:
\begin{equation}
	K(u,u) = \frac{1}{4\pi^2 l^2} \left[ -P_+ (\gamma^0\gamma^5 \mu_5 + \gamma^3\kappa) + 
	\gamma^0\gamma^5 \mu_5 \;f\left(\frac{2|eB|}{\mu_5^2}\right)
	+ \gamma^3\frac{a\kappa\Lambda^2}{2|eB|} \right],
\end{equation}
where $a=2(\sqrt{2}-1)$ and $f(z)=\sum_0^{[1/z]}\sqrt{1-nz}$ is a function that behaves like $2/(3z)$ at small $z$. Equations for the $\kappa$ and $\mu_5$ then have the following form:
\begin{equation}
	\begin{aligned}
		\kappa =& \frac{G_{\rm int}}{4\pi^2 l^2} \left[-s_\bot \mu_5 - \kappa + \frac{a\kappa\Lambda^2}{|eB|} \right], \\
		\mu_5-\mu_5^0 =& \frac{G_{\rm int}}{4\pi^2l^2} \left[\mu_5+s_\bot\kappa - 2\mu_5 \;f\left(\frac{2|eB|}{\mu_5^2}\right)  \right] .
	\end{aligned}
\end{equation}
By using the iteration method, we find the following approximate solution:
\begin{equation}
	\begin{aligned}
		\kappa =& - \frac{g \mu_5 s_\bot / (\Lambda l)^2}{1 - a g} ,\\
		\mu_5 =& \frac{\mu_5^0+gs_\bot\kappa/(\Lambda l)^2}{1+(2f_0-1)g/(\Lambda l)^2},
	\end{aligned}
\end{equation}
where we introduced the dimensionless coupling constant $g\equiv G_{\rm int}\Lambda^2/(4\pi^2)$ and $f_0\equiv f\left(2|eB|/(\mu_5^0)^2\right)$.

\subsection{Proper-time regularization}
By making use of the proper-time regularization, the integral in Eq.~(\ref{eq:full-propagator-in-one-point}) can be rewritten as follows:
\begin{equation}
	\int dk_3  \frac{k_3-\kappa}{\sqrt{(k_3-\kappa)^2 + 2n|eB|}} = \int dk_3 \int_0^\infty \frac{d\alpha}{\sqrt{\alpha\pi}} (k_3-\kappa)e^{-\alpha[(k_3-\kappa)^2 + 2n|eB|]} = 0,
\end{equation}
where the last equality comes from symmetry. We then can rewrite Eq.~(\ref{eq:full-propagator-in-one-point}) as
\begin{equation}
	K(u,u) = 
	\frac{1}{4\pi^2 l^2} \left[ -P_+  + f\left(\frac{2|eB|}{\mu_5^2}\right) \right] \gamma^0\gamma^5 \mu_5,
\end{equation}
and after repeating the same steps we will get:
\begin{equation}
	\begin{aligned}
		\kappa =& - g \mu_5 s_\bot / (\Lambda l)^2, \\
		\mu_5 \approx& \frac{\mu_5^0}{1+(2f_0-1)g/(\Lambda l)^2}.
	\end{aligned}
\end{equation}

It is also interesting to compute the expectation value of the third component of the current $\langle j_z \rangle = \langle \gamma^3 G \rangle$ and its first order modification due to interaction:
\begin{equation}
\label{eq:current_NJL}
	\langle j_z \rangle = -\frac{2}{G_{\rm int}} \kappa = 
		\begin{dcases}
		\frac{2\mu_5^0 s_\bot}{4\pi^2 l^2} \left[1 + g\left(a+\frac{2f_0-1}{(\Lambda l)^2}\right) + O(g^2) \right] = \frac{2\mu_5^0 s_\bot}{4\pi^2 l^2} - \frac{2\kappa}{4\pi^2 l^2}\left(a\Lambda^2 l^2+2f_0-1\right) + O(\kappa^2)
		\\
		\frac{2 \mu_5^0 s_\bot}{4\pi^2 l^2} \left[1 + g\frac{2f_0-1}{(\Lambda l)^2} + O(g^2) \right] = \frac{2\mu_5^0 s_\bot}{4\pi^2 l^2} - \frac{2\kappa}{4\pi^2 l^2}\left(2f_0-1\right) + O(\kappa^2),
		\end{dcases}
\end{equation}
where the two lines give the results for the cutoff and proper-time regularizations, respectively.

As we can see, both schemes successfully reproduce the usual quantum anomaly result for the current. However, they give different 
interaction induced corrections. This may be due to the fact that the anomaly in the nonrenormalizable model at hand is especially sensitive to high energy states, which are beyond the range of validity of our low-energy 
model. The momentum shift parameter $\kappa$ also appears to be independent of momenta and Landau level index (which is natural for point-like 
interactions) and thus formally can be removed from the effective Lagrangian by a gauge transformation. We expect it to become momentum-dependent in a more realistic QED theory.

\section{Chiral parameter in QED: weak field limit}
\label{sec:QED}

Let us now consider more realistic case of the QED theory in the chiral limit, as was done in Ref.~\cite{Gorbar:2013upa,chiral-shift-QED}. The system 
consists of massless fermions with a nonzero chiral chemical potential $\mu_5$ in the external electromagnetic field $B$. As usual we take the direction of the $z$ axis to be parallel to $B$. The Lagrangian has the form:
\begin{equation}
	\mathcal{L} = -\frac{1}{4} F^{\mu\nu}F_{\mu\nu} + \bar{\psi} (i \gamma^\mu \mathcal{D}_\mu + \mu_5 \gamma^5 \gamma^0) \psi
	+\mbox{(counter terms)},
\end{equation}
where $\mathcal{D}_\mu = \partial_\mu - ieA^{\rm ext}_\mu - ieA_\mu$ and the vector potential consists of the background magnetic field potential in the Landau 
gauge $A_{\rm ext}^\mu=(0,0,xB,0)$ and the quantum part of the electromagnetic field $A$. The field strength tensor $F^{\mu\nu}$ is defined by $F^{\mu\nu}=\partial_\mu A_\nu - \partial_\nu A_\mu$. We also added the counter terms to the Lagrangian to regularize the divergences in the loop diagrams of the self-energy.

There are two symmetry breaking effects present in the system. The parity symmetry interchanges left- and right-handed fermions and, thus, is broken when a nonzero chiral chemical potential $\mu_5$ is present. The time reflection symmetry is, in turn, broken by the background magnetic field $B$. The latter 
also breaks the $O(3)$ rotational group to $O(2)$ rotations around the $z$ axis. Because of these broken discrete symmetries, a plethora of $P$ and $T$ breaking matrix structures are allowed in the full propagator.

The free propagator of a fermion in a uniform magnetic field can be factorized into a translation invariant part $\tilde{S}$ 
and the Schwinger phase $\Phi(x,y)=-eB(x_1+y_1)(x_2-y_2)/2$, similar to Ref.~\cite{Gorbar:2013upa}, i.e.,
\begin{equation}
	S(x,y) = e^{i \Phi(x,y)} \tilde{S}(x-y).
\label{eq:propagator-factorization}
\end{equation}
To reduce the notation burden without loosing generality we assume that $\sign(eB)=+1$. Then the translation invariant part of the propagator in the momentum space has the form:
\begin{equation}
\begin{gathered}
	\tilde{S}(k) = 2i e^{-\bs{k}_\bot^2 l^2} \sum_{s_5} \sum_{n=0}^{\infty} \frac{(-1)^n D_n(k) }{(k_0+s_5\mu_5)^2-k_3^2-2n|eB|} P_{s_5}, \\
	D_n(k) = (k_0 \gamma^0-k_3\gamma^3+s_5\mu_5\gamma^0) [P_- L_n(2\bs{k}_\bot^2 l^2) - P_+L_{n-1}(2\bs{k}_\bot^2 l^2)] + 2(\bs{k}_\bot \cdot \bs{\gamma}_\bot) L^1_{n-1}(2\bs{k}_\bot^2 l^2),
\end{gathered}
\label{eq:propagator-Landau-expanded}
\end{equation}
where we introduced the spin and chirality projectors: $P_\pm=(1\pm i\gamma^1\gamma^2)/2$ and  $P_{s_5}=(1+s_5\gamma^5)/2$, respectively.

In a general case, the analysis can be performed using the generalized Landau-level representation. Here, for simplicity, we will consider the problem in the weak field limit. To the leading linear order in the magnetic field, the fermion propagator takes the following form:
\begin{equation}
	\tilde{S}(k) = \bar{S}^{(0)}(k) +\bar{S}^{(1)}(k) +\cdots,
\end{equation}
where
\begin{equation}
	\tilde{S}^{(0)}(k) = i \sum{s_5} \frac{(k_0+s_5\mu_5)\gamma^0- \mathbf{k}\cdot\bm{\gamma}}
	{(k_0+s_5\mu_5+i\epsilon\, \mbox{sign}(k_0))^2-\mathbf{k}^2} P_{s_5}
\label{free-term}
\end{equation}
and 
\begin{equation}
	\tilde{S}^{(1)}(k) =  - \gamma^1\gamma^2 eB \sum_{s_5} \frac{(k_0+s_5\mu_5)\gamma^{0}-k_3 \gamma^3}
	{\left[ (k_0+s_5\mu_5+i\epsilon\, \mbox{sign}(k_0))^2-\mathbf{k}^2 \right]^2 }.
\label{linear-term}
\end{equation}

The self-energy to the lowest order in $\alpha$ has the form:
\begin{equation}
	\Sigma(p) = - 4\pi\alpha \int \frac{d^4k}{(2\pi)^4} \gamma_\mu S(k) \gamma^\mu \frac{1}{(k-p)^2},
\label{eq:self-energy_general}
\end{equation}
which can be written in the same form as the fermion propagator in Eq.~(\ref{eq:propagator-factorization}), i.e., the product of the Schwinger phase factor and a translation invariant part $\tilde{\Sigma}$. Following Ref.~\cite{Gorbar:2013upa} we then divide the self-energy into the vacuum and material parts for the computational purposes. In the zeroth order of $B$ the vacuum part is responsible for wavefunction renormalization and in Pauli-Villars regularization is given by:
\begin{equation}
\begin{aligned}
	\tilde{\Sigma}^{(0)}_{\rm vac}(p) &= - \frac{\alpha}{2\pi} \sum_{s_5} [(p_0-s_5\mu_5)\gamma^0 - \bs{p}\cdot\bs{\gamma}] \int_0^1 dx (1-x)\ln\frac{\Lambda^2}{x[\bs{p}^2-(p_0-s_5\mu_5)^2]} P_{s_5} 
	\\
	&= \frac{\alpha}{4\pi} \sum_{s_5} [(p_0-s_5\mu_5)\gamma^0 - \bs{p}\cdot\bs{\gamma}] \left(\frac{3}{2}+\ln\frac{\Lambda^2}{\bs{p}^2-(p_0-s_5\mu_5)^2}\right)P_{s_5}.
\end{aligned}
\label{eq:self-energy_vac_0}
\end{equation}

This self-energy is identical to the usual QED self-energy after substitution $(p_0-\mu_5\gamma^5)\rightarrow p_0$. This means that we can use the same reasoning and incorporate the divergence into wavefunction renormalization constant $Z_2=1+\delta_2$, where
\begin{equation}
	\delta_2 = \left.\frac{d\Sigma}{d\slashed{P}}\right\vert_{\slashed{P}=\lambda}=\frac{\alpha}{4\pi} \left(\ln\frac{\Lambda^2}{-\lambda^2} - \frac{1}{2} \right),
\end{equation}
and we defined $P=(p_0-\mu_5\gamma^5,\bs{p})$. The usual choice for the pole is $\slashed{P}=m$, but in the case of the massless fermions we need to choose an arbitrary renormalization scale $\lambda$ instead. For concreteness, in our calculations below, we will fix $\lambda$ to be equal to $\mu_5$. Note, that the self-energy in Eq.~(\ref{eq:self-energy_vac_0}) is zero for $\slashed{P}=0$, and this means that the radiative corrections do not generate mass. The finite part equals:
\begin{equation}
\begin{aligned}
	\tilde{\Sigma}^{(0)}_{\rm vac}(p) - \delta_2 \slashed{P} &= \sum_{s_5} [(p_0-s_5\mu_5)\gamma^0 - \bs{p}\cdot\bs{\gamma}] \sigma_{\rm vac}(p) P_{s_5}, \\
	\sigma_{\rm vac}(p)&\equiv  \frac{\alpha}{4\pi}  \left(2+\ln\frac{\lambda^2}{(p_0-s_5\mu_5)^2-\bs{p}^2}\right).
\end{aligned}
\label{eq:self-energy_vac_0}
\end{equation}
The material part in the zeroth order of $B$ is given by:
\begin{equation}
\begin{aligned}
	\tilde{\Sigma}^{(0)}_{\rm mat}(p) =& \frac{\alpha\gamma^0}{4\pi\bs{p}^2} \Bigg( (p_0^2-\mu_5^2-\bs{p}^2)|\mu_5| 
	- \frac{\big(p_0^2-(|\mu_5|-|\bs{p}|)^2\big)\big(p_0^2-(|\mu_5|+|\bs{p}|)^2\big)}{4|\bs{p}|} \ln\frac{p_0^2-(|\mu_5|-|\bs{p}|)^2}{p_0^2-(|\mu_5|+|\bs{p}|)^2} \Bigg)\\
	+ &\sum_{s_5} [(p_0-s_5\mu_5)\gamma^0 - \bs{p}\cdot\bs{\gamma}] \sigma_{\rm mat}(p) P_{s_5}, \\
	\sigma_{\rm mat}(p)\equiv&  \frac{\alpha}{16\pi|\bs{p}|^3} \left( 4|\mu_5\bs{p}|(p_0+s_5\mu_5)
	- 2|\bs{p}|^3\,\ln\frac{(p_0-|\mu_5|)^2-\bs{p}^2}{(p_0+|\mu_5|)^2-\bs{p}^2}  \right.
	\\
	&\hspace{0.25\textwidth}\left. -[2p_0\bs{p}^2-(p_0+s_5\mu_5)(p_0^2-\bs{p}^2-\mu_5^2)]\,\ln\frac{p_0^2-(|\bs{p}|-|\mu_5|)^2}{p_0^2-(|\bs{p}|+|\mu_5|)^2} \right),
\end{aligned}
\label{eq:self-energy_mat_0}
\end{equation}
where we defined the vacuum and material part of the momentum renormalization $\sigma_{\rm vac}(p)$ and $\sigma_{\rm mat}(p)$. In the first order of $B$ the self-energy has the form:
\begin{equation}
	\tilde{\Sigma}^{(1)}_{\rm vac}(p) = \frac{\alpha eB}{2\pi} i\gamma^1\gamma^2 \sum_{s_5}  \frac{(p_0-s_5\mu_5)\gamma^0 - p_3 \gamma^3 }{(p_0-s_5\mu_5)^2 - \bs{p}^2}P_{s_5}
\label{eq:self-energy_vac_1}
\end{equation}
and
\begin{equation}
\begin{aligned}
	\tilde{\Sigma}^{(1)}_{\rm mat}(p) =& \frac{\alpha eB}{8\pi |\bs{p}|} i\gamma^1 \gamma^2 \sum_{s_5}
	\Bigg\{\gamma^0 \left( 4\frac{s_5\mu_5|\bs{p}|}{(p_0-s_5\mu_5)^2-\bs{p}^2} + s_5\sign\mu_5\,\ln\frac{p_0^2-(|\mu_5|-|\bs{p}|)^2}{p_0^2-(|\mu_5|+|\bs{p}|)^2} \right) 
	\\
	&\hspace{0.15\textwidth}+ \frac{p_3\gamma^3}{|\bs{p}|} \left(4 \frac{\mu_5^2-s_5\mu_5p_0}{(p_0-s_5\mu_5)^2-\bs{p}^2} + s_5\sign\mu_5\,\frac{s_5\mu_5+p_0}{|\bs{p}|} \ln\frac{p_0^2-(|\mu_5|-|\bs{p}|)^2}{p_0^2-(|\mu_5|+|\bs{p}|)^2} \right) \Bigg\} P_{s_5},
\end{aligned}
\label{eq:self-energy_mat_1}
\end{equation}
The only terms present in the self-energy are proportional to $\gamma^0$, $\gamma^0\gamma^5$, $\gamma^3$ and $\gamma^3\gamma^5$ and are responsible for the ordinary and chiral chemical potentials as well as the momentum and chiral shift parameters respectively.

\subsection{Modification of the dispersion relation}

Using the self-energy we can find the poles of the full fermion propagator approximately from the following equation, used for the case of the usual chemical potential in Ref.~\cite{chiral-shift-QED}:
\begin{equation}
	\det(iS^{-1}-\tilde{\Sigma})=0,
\label{eq:poles_eqn}
\end{equation}
where the self-energy $\tilde{\Sigma}$ equals the sum of all previous calculated self-energy expressions in Eq.~(\ref{eq:self-energy_vac_0})-(\ref{eq:self-energy_mat_1}):
\begin{equation}
	\tilde{\Sigma} = \sum_{s_5} [(p_0-s_5\mu_5)\gamma^0 - \bs{p}\cdot\bs{\gamma}] \sigma(p) P_{s_5} + 
	\Sigma_0 \gamma^0 + \Sigma_3 \gamma^3 + \Sigma_{05} \gamma^0\gamma^5 + \Sigma_{35} \gamma^3\gamma^5,
\label{eq:self-energy}
\end{equation}
and we denoted the total momentum renormalization $\sigma(p)\equiv\sigma_{\rm vac}(p)+\sigma_{\rm mat}(p)$.

Since the fermions are massless, the dispersion relation separates for the left- and right-handed fermions and we end up with the determinant of a block matrix, where each block represents corresponding chirality:
\begin{equation}
\begin{aligned}
	\text{det}\left( \sum_{s_5} \left\{(1-\sigma(p))[(p_0-s_5\mu_5)\gamma^0 - \bs{p}\cdot\bs{\gamma}] - (\Sigma_0+s_5\Sigma_{05})\gamma^0 - (\Sigma_3+s_5\Sigma_{35})\gamma^3\right\} P_{s_5} \right) = 0,
\end{aligned}
\label{eq:dispersion-determinant}
\end{equation}
where $s_5=+1$ and $s_5=-1$ correspond to the right- and left-handed particles respectively. Note, that the corrections to the unperturbed dispersion relation have the order of $\alpha\ln\alpha$, whereas the momentum renormalization factor $(1-\sigma(p))$ will be relevant only in the $(\alpha\ln\alpha)^2$ order. We will omit that factor in what follows. Note, however, that we took it into account in the numerical calculations. In the lowest order the dispersion relations for fermions and anti-fermions of both chiralities have the form:
\begin{equation}
\begin{aligned}
	&\def\arraystretch{1.5}
	\begin{array}{r}
		p_0 = \mu_5 + (\Sigma_0+\Sigma_{05}) + \sqrt{\bs{p}_\bot^2 + (p_3+\Sigma_3+\Sigma_{35})^2} \\
		p_0 = \mu_5 + (\Sigma_0+\Sigma_{05}) - \sqrt{\bs{p}_\bot^2 + (p_3+\Sigma_3+\Sigma_{35})^2}
	\end{array}
	&\qquad-\text{right},
	\\
	&\def\arraystretch{1.5}
	\begin{array}{r}
		p_0 = -\mu_5 + (\Sigma_0-\Sigma_{05}) + \sqrt{\bs{p}_\bot^2 + (p_3+\Sigma_3-\Sigma_{35})^2} \\
		p_0 = -\mu_5 + (\Sigma_0-\Sigma_{05}) - \sqrt{\bs{p}_\bot^2 + (p_3+\Sigma_3-\Sigma_{35})^2}
	\end{array}
	&\qquad-\text{left},
\end{aligned}
\label{eq:dispersion_equations}
\end{equation}

\begin{figure}
	\includegraphics[width=0.46\textwidth]{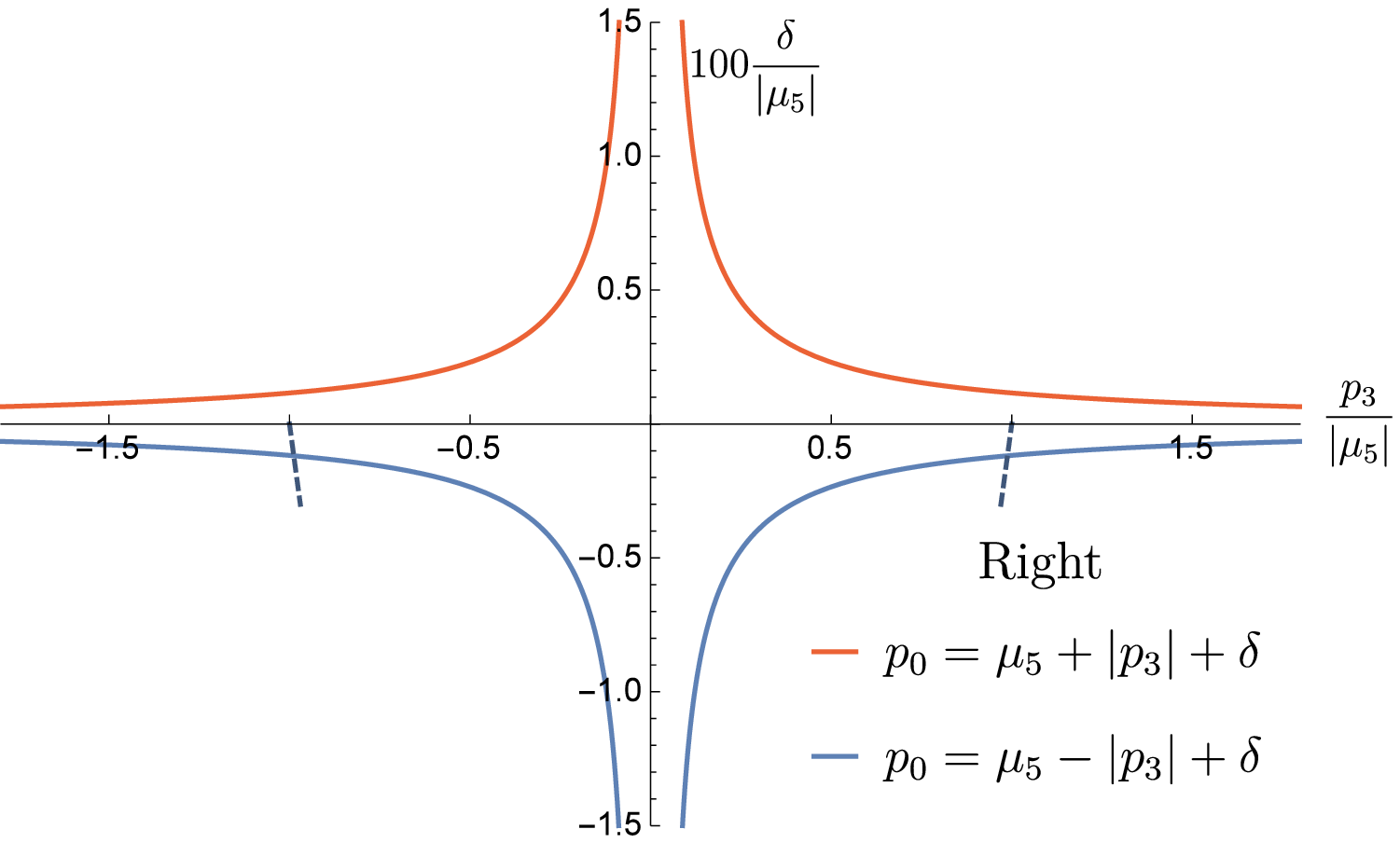}\qquad
	\includegraphics[width=0.46\textwidth]{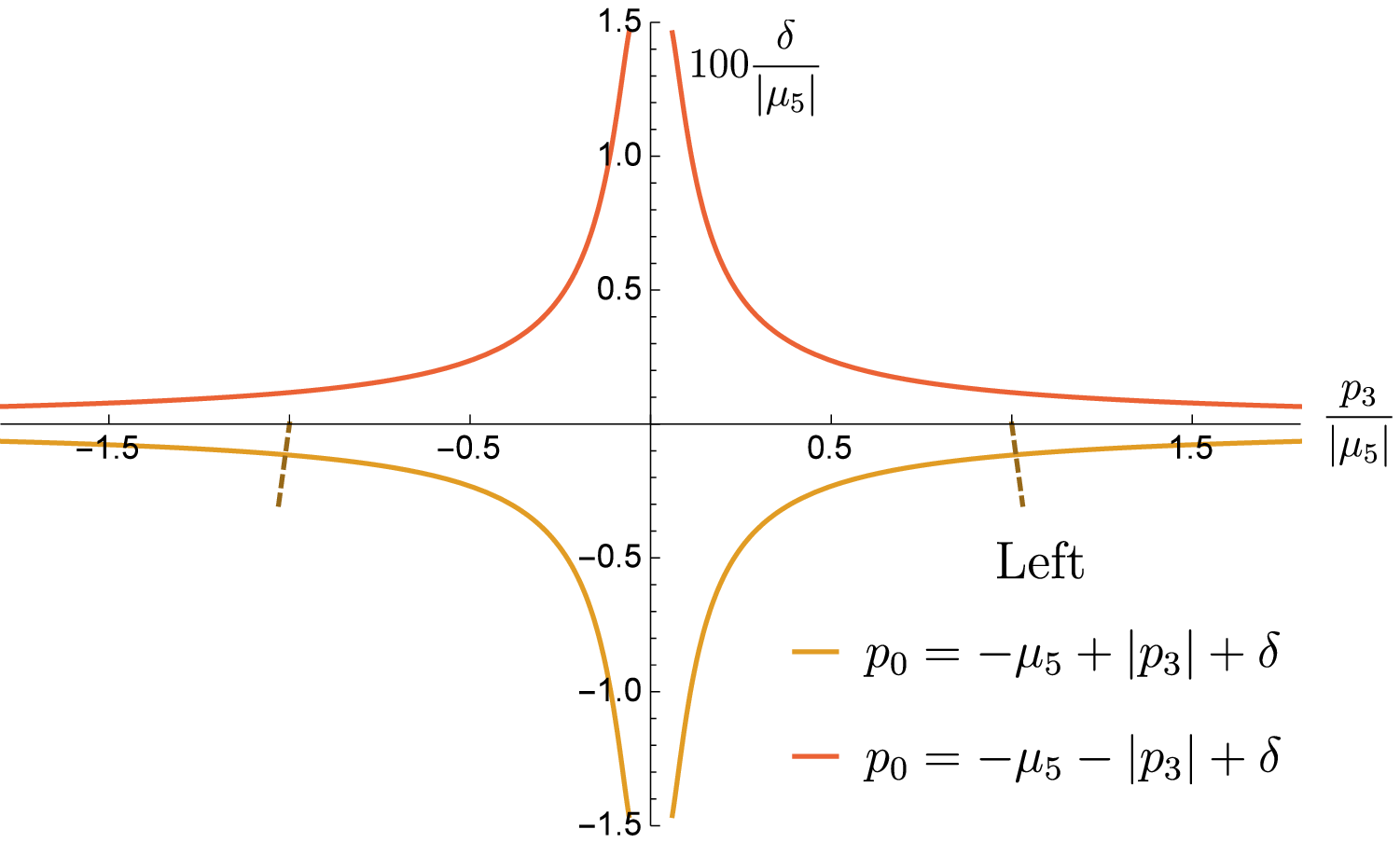}\\
	\includegraphics[width=0.46\textwidth]{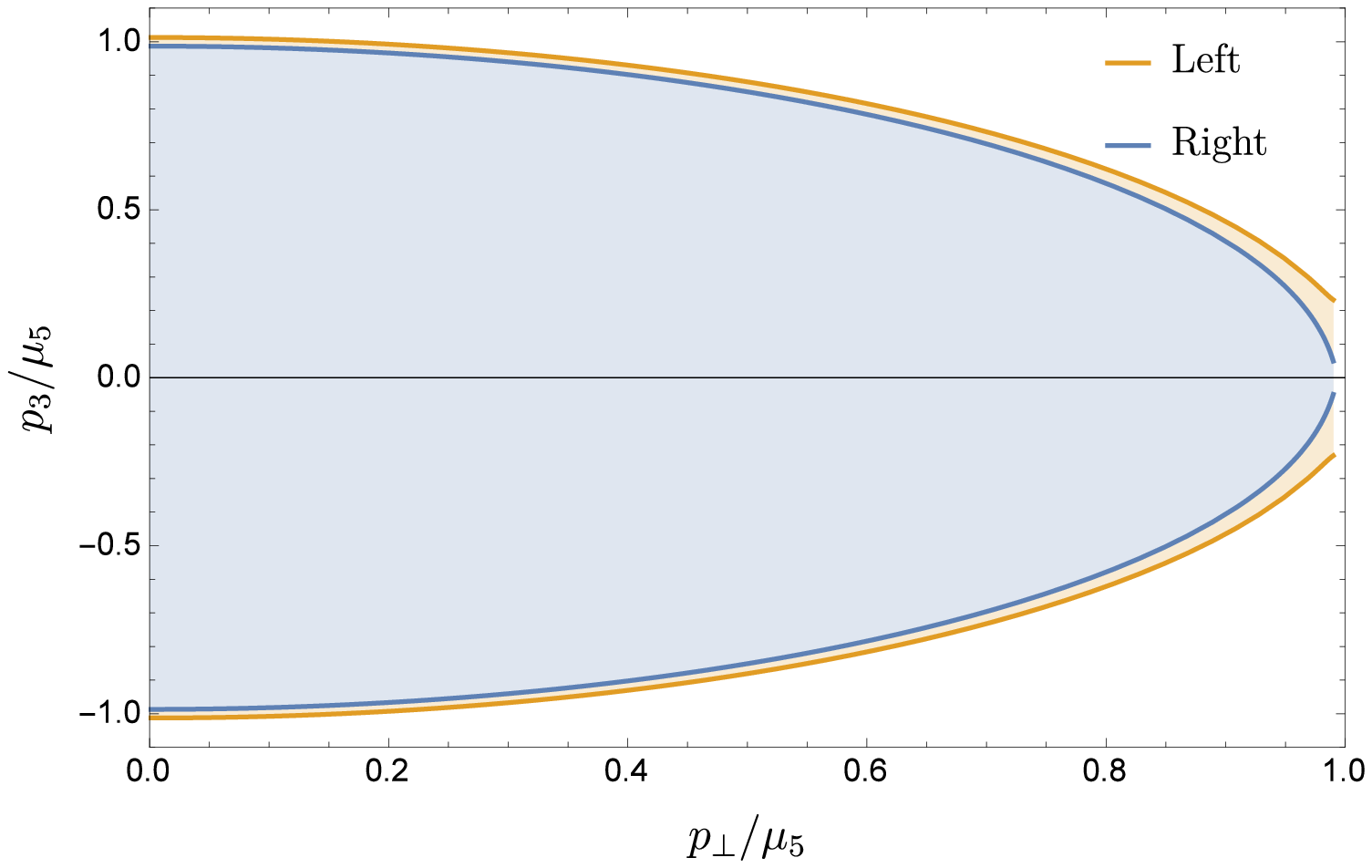}
	\caption{\textbf{Top:} The modification of the dispersion relations $\delta$ and the Fermi-surfaces for the right- and left-handed fermions for $p_\bot=0$, $\mu_5>0$, $eB = 0.01\,\mu_5^2$ and $\alpha=1/137$. The difference between the branches of different chirality on the bottom panel is enhanced to show the qualitative positioning, on the genuine graph the branches are indistinguishable.}
\label{fig:fig}
\end{figure}

The self-energy structures in Eq.~(\ref{eq:self-energy}) have the form:
\begin{equation}
\begin{aligned}
	\Sigma_0 =& \frac{\alpha(p_0^2-\mu_5^2-\bs{p}^2)|\mu_5|}{4\pi\bs{p}^2} 
	- \frac{\alpha\big(p_0^2-(|\mu_5|-|\bs{p}|)^2\big)\big(p_0^2-(|\mu_5|+|\bs{p}|)^2\big)}{16\pi|\bs{p}|^3} \ln\frac{p_0^2-(|\mu_5|-|\bs{p}|)^2}{p_0^2-(|\mu_5|+|\bs{p}|)^2} 
	\\
	&\quad-\frac{\alpha eB}{2\pi} \frac{p_3}{\bs{p}^2} \frac{\mu_5p_0(p_0^2+\bs{p}^2-\mu_5^2) }{\bs{p}^4-2\bs{p}^2(p_0^2+\mu_5^2)+(p_0^2-\mu_5^2)^2} 
	+ \frac{\alpha eB\sign\mu_5}{8\pi} \frac{p_0p_3}{|\bs{p}|^3} \ln\frac{p_0^2-(|\mu_5|-|\bs{p}|)^2}{p_0^2-(|\mu_5|+|\bs{p}|)^2},
	\\[1.0ex]
	\Sigma_{05} =& -\frac{\alpha eB}{2\pi} \frac{p_3}{\bs{p}^2} \frac{p_0^2(\bs{p}^2+\mu_5^2)-(\bs{p}^2-\mu_5^2)^2 }{\bs{p}^4-2\bs{p}^2(p_0^2+\mu_5^2)+(p_0^2-\mu_5^2)^2} 
	+ \frac{\alpha eB}{8\pi} \frac{|\mu_5|p_3}{|\bs{p}|^3} \ln\frac{p_0^2-(|\mu_5|-|\bs{p}|)^2}{p_0^2-(|\mu_5|+|\bs{p}|)^2},
	\\[1.0ex]
	\Sigma_3 =& \frac{\alpha eB}{2\pi} \frac{2\mu_5p_0^2 }{\bs{p}^4-2\bs{p}^2(p_0^2+\mu_5^2)+(p_0^2-\mu_5^2)^2} 
	+ \frac{\alpha eB\sign\mu_5}{8\pi|\bs{p}|} \ln\frac{p_0^2-(|\mu_5|-|\bs{p}|)^2}{p_0^2-(|\mu_5|+|\bs{p}|)^2},
	\\[1.0ex]
	\Sigma_{35} =& \frac{\alpha eB}{2\pi} \frac{p_0(p_0^2-\bs{p}^2+\mu_5^2) }{\bs{p}^4-2\bs{p}^2(p_0^2+\mu_5^2)+(p_0^2-\mu_5^2)^2}.
\end{aligned}
\label{eq:self-energy-expressions}
\end{equation}
Note, that only the term with $\gamma^0$ gets modified in the absence of the magnetic field. The unperturbed dispersion relation in the momentum space for plasma with the chiral chemical potential consists of two cones shifted in $p_0$ coordinate up and down by $\mu_5$ from the origin, where each cone represents different chirality, i.e. $p^{\rm (cone)}_0=\pm(\mu_5\pm |\bs{p}|)$. This dispersion will be modified by the interaction induced self-energy corrections and so we look for the solutions of Eq.~(\ref{eq:dispersion_equations}) in the form $p_0=p^{\rm (cone)}_0+\delta$. We are mostly interested in the behavior along $B$, so we put $p_\bot=0$. The dispersion relation modifications for the right- and left-handed fermions given in the top panel Fig.~\ref{fig:fig} were obtained numerically for $\mu_5>0, eB = 0.01 \mu_5^2$ and $\alpha=1/137$. As can be seen from the graphs the absolute value of $\delta$ gradually increases towards the cones vertices, until our one-loop calculations break up. The behavior of such plasma is heavily dependent on its Fermi-surface, which is determined as an intersection of the dispersion with the condition $p_0=0=\pm(\mu_5-|p_3|)+\delta$ shown on the graphs with the dashed lines. As can be seen from the graphs the Fermi-surface for the right-handed particles shrinks, whereas it expands for the left-handed particles.

To investigate the Fermi-surfaces modification further we solve Eq.~(\ref{eq:dispersion_equations}) for $p_0=0$. The result is presented on the bottom panel of Fig.~\ref{fig:fig}. The Fermi-surfaces are modified in the lowest order by the two effects: the uniform shrinking (expansion) for the right- (left-)handed particles, as qualitatively predicted in the dispersion analysis, and shift in the positive direction of $p_3$. The first effect is a result of the zeroth order in $B$ self-energy contributions (\ref{eq:self-energy_vac_0})-(\ref{eq:self-energy_mat_0}). The modification has a different sign for the left- and right-handed particles and is of order of $\alpha/(2\pi\sqrt{1-p_\bot^2/\mu_5^2})$. Note, however, that for a given chirality it is the same for the opposite $p_3$ and in this way does not generate current. The second effect is a shift in the positive direction of $p_3$ induced by the first order of $B$ self-energy corrections (\ref{eq:self-energy_vac_1})-(\ref{eq:self-energy_mat_1}). The main contribution of this shift is of order of $\alpha eB/(\pi\mu_5)$ and is independent of $p_\bot$, as well as particles chirality or the $p_3$ sign. It is several orders weaker than the first effect due to the dependence on the small magnetic field, however it is responsible for the current generation along the $z$ direction with particles of both chiralities contribute. This confirms the predictions of the NJL analysis done in the Sec.~\ref{sec:simpl} of this paper that the magnetic field leads to the current generation in the systems with particles of different chirality imbalance. Note, that the induced Fermi-surface shift is at least several orders smaller than the one produced by the ordinary chemical potential $\mu$ considered in Ref.~\cite{chiral-shift-2}.

\section{Conclusion}
In this paper we considered a cold relativistic plasma with nonzero chiral chemical potential $\mu_5$ in the background uniform magnetic field $B$. Our analytic analysis in the NJL-model showed that a nonzero correction to the electric current is induced due to interactions. This is a manifestation of radiative correction to the chiral magnetic effect. As was argued in the dispersion relations analysis, both left-handed and right-handed states near the Fermi energy are shifted towards the positive $p_3$ direction, which implies an electric current generation in the positive $z$ direction.

It can be seen from the self-energy expressions in Eq.~(\ref{eq:self-energy-expressions}) that four parameters proportional to the magnetic field are generated in the system. The chemical potential term proportional to $\gamma^0$ is sensitive to the sign of $\mu_5$ and $p_3$. This can modify the current along the $z$ direction. The momentum shift term proportional to $\gamma^3$ is analogous to the $\kappa$ term in the NJL analysis performed in the first part of this paper. It is dependent on the sign of $\mu_5$ and is responsible for the current generation along the magnetic field. The chiral momentum shift parameter proportional to $\gamma^3\gamma^5$ is independent of the signs of $\mu_5$ and $p_3$, and presumably can lead to the chiral current generation in the system. The last term proportional to $\gamma^0\gamma^5$ is the chiral chemical potential modification and plays the same role as $\mu_5-\mu_5^0$ in the NJL analysis.

Our results in QED not only reconfirm the NJL model predictions regarding the momentum shift generation, but also reveal the generation of the chemical potential and chiral momentum shift. Unlike the usual chiral magnetic effect (CME), which originates from the fermions on the LLL \cite{Fukushima:2012vr}, these corrections are interaction induced and involve higher Landau levels as well. This can lead to the observable effects in systems with chiral fermions, such as the early Universe plasma, heavy-ion collisions or Weyl semimetals. Though the analysis was performed in a zero temperature approximation previous studies of the analogous systems (see Ref.~\cite{chiral-shift-2}) showed a weak dependence on the temperature, which suggests that similar effects can survive even in high temperature.

As compared to the case of the ordinary chemical potential $\mu\neq0$ in a magnetic field considered in Ref.~\cite{chiral-shift-QED} the self-energy has the additional chiral momentum and chemical potential parameters, which can be attributed to the fact that more symmetries are broken in the present case. This means that nothing forbids $\bs{j}$ and $\bs{j}_5$ to be both induced by interactions.

\begin{acknowledgments}
The author would like to thank I.A. Shovkovy and E.V. Gorbar for proposing this project, as well as for useful discussions.
\end{acknowledgments}


\end{document}